# Solution-Processable Graphene Oxide as an Efficient Hole Injection Layer for High Luminance Organic Light-Emitting Diodes

Shengwei Shi,*[ab] Veera Sadhu,[a] Reda Moubah,[c] Guy Schmerber,[d] Qinye Bao[b] and S. Ravi P. Silva*[a]

[a] Nanoelectronics Center, Advanced Technology Institute, University of Surrey, Guildford, GU2 7XH, Surrey, United Kingdom. E-mail: s.silva@surrey.ac.uk

[b] Department of Physics, Chemistry and Biology, Linköping University, S-58183 Linköping, Sweden. E-mail: shesh@ifm.liu.se

[c] Department of Physics and Astronomy, Uppsala University, S-75120 Uppsala, Sweden.

[d] IPCMS, Université de Strasbourg, 23 rue du Loess, F-67034 Strasbourg, France.

**ABSTRACT:**

   The application of solution-processable graphene oxide (GO) as hole injection layer in organic light-emitting diodes (OLEDs) is demonstrated. High luminance of over 53,000 cd m$^{-2}$ is obtained at only 10 V. The results will unlock a route of applying GO in flexible OLEDs and other electrode applications.

Over the last two decades significant advances have been made in organic light-emitting diodes (OLEDs) due to their applications in flat panel displays and solid state lighting.[1] It is well known that the performance of OLEDs is largely dominated by charge injection from electrodes.[2] Indium tin oxide (ITO) is the most widely used electrode for OLEDs because of its high optical transparency and good conductivity, but the expense of indium and its brittleness limits its usage on flexible substrates. The surface electronic properties of ITO are still less than ideal, in particular, a large hole-injection barrier is found at the ITO/organic interface due to the mismatch between its work function (WF) and energy levels of the organics.[3,4] Various hole-injection layers and strategies have been introduced to improve the hole injection and its path towards the radiative recombination zone in the emission layer. Poly(3,4-ethylenedioxythiophene):poly(styrene sulfonate) (PEDOT:PSS) is widely used in polymer light-emitting diodes (PLEDs) as one such hole transport layer. However, PEDOT:PSS can cause degradation of the ITO due to its acidic nature, particularly in the presence of moisture.[5] Transition metal oxides (TMOs) have also been used to reduce hole-injection barrier, but high temperature from the common vacuum deposition can potentially be detrimental to device performance,[6] and there are some reports on solution processed TMOs, however, the processes become complicated because of the precursor synthesis and high temperature annealing.[7] P-doped organic layers are reported to effectively enhance hole injection in OLEDs, but doping technology is complicated with much adjusting of the doping ratio.[8] Other strategies for hole injection has been attempted based on nano-carbon electrodes with limited success.[9-11] In comparison, a solution-processable and efficient hole-injecting material such as graphene oxide (GO) that can replace PEDOT:PSS is highly desirable, and has the potential of cheap and flexible ITO-free electrodes in organic optoelectronics.

Despite its great potential as a transparent conductor, the application of graphene as the anode in organic optoelectronic has been limited because of its relatively low WF and high sheet resistance compared with ITO.[12] Graphene-based OLEDs have shown poorer device efficiencies with higher operating voltages than ITO-based devices,[13,14] although additional hole injection layers can change this process with high efficiency OLEDs.[12] GO is a graphene sheet functionalized with oxygen groups in the form of epoxy and hydroxyl groups on the basal plane and various other types at the edges.[15] It is generally used as a precursor for graphene,[16] and shows a great potential for use in devices such as organic solar cells.[17,18] Comparing with graphene, GO has a better solution-processable, chemically tunable structure, with reproducible properties and simple device fabrication processes.[19] Currently, there are some reports on GO in OLEDs.[20-24] Functionalized GO is used as hole injection layer in doped OLEDs, showing better performance over the reference device in terms of turn-on voltage, current and power efficiency, for example, the current efficiency can be enhanced by 150%, while the highest luminances are obtained at the voltage over 17 V.[20] A highly controllable thin film of reduced GO is reported to act as anode in OLEDs, but the device performance still cannot compete with the control devices based on ITO.[21] Reduced GO has also been used as a cathode in an inverted PLEDs and light-emitting electrochemical cell because of its tunable work function.[22,23] Only very recently was GO proposed to be used as hole injection layers in PLEDs to show much better device performance than the control ITO device.[24]

Here, we report a non-doped, high luminance, fluorescent OLED in a simple device structure with solution-processable GO as hole injection layer. All devices are fabricated on pre-patterned ITO with N,N'-bis(3-methylphenyl)-N,N'-diphenyl-1,1'-biphenyl-4,4'-diamine (TPD) as the hole transport layer and Tris(8-hydroxyquinolinato)aluminum ($Alq_3$) as the electron transport and emission layer. The device shows a highest luminance of over 53,000 cd m$^{-2}$ at 10 V, and it shows a record luminance of 40,785 cd m$^{-2}$ at only 8.8 V, when a high concentration GO is used. There is a sharp increase of nearly 30 times as compared to the reference device.

To characterize the fundamental structure of GO and evaluate its potential application in OLEDs, we conduct several analyses such as atomic force microscopy (AFM), Fourier transform infrared spectroscopy (FT-IR), X-ray photoemission spectroscopy (XPS) and ultraviolet visible spectroscopy (UV-vis). The films of GO readily form on various substrates (e.g., glass, silicon and ITO) by spin-coating of its solution in deionized water (DI). The topography of GO film from the solution with a concentration of 0.1 mg ml$^{-1}$ on silicon substrate is checked by AFM in a tapping mode (Fig.1 (a)). The average Ra roughness is about 0.58 nm. The height of the film is between 1 to 1.5 nm, since the acceptable height range for modified single-layer graphene is around 1 nm, GO shows a monolayer or slightly thicker sheet on silicon. Because of the low concentration and two-dimensional nature, a large surface of GO film with good uniformity is obtained for device fabrication. As shown in Fig.1 (b), the FTIR spectra of GO and the raw graphite flakes reveal the presence of oxygen-containing functional groups in GO. The peaks at 972-1040 cm$^{-1}$, 1160-1380 cm$^{-1}$, and 1630 cm$^{-1}$ all correspond to C-O-C stretching vibrations, C-OH stretching, and C-C stretching mode of SP$^2$ carbon skeletal network, respectively. The peak at 1725 cm$^{-1}$ corresponds to C-O stretching vibrations of the COOH groups and the peak at 3000-3550 cm$^{-1}$ corresponds to O-H stretching vibrations. These functional groups make GO highly hydrophilic and render it dispersible compared to raw graphite flakes. XPS characterization also confirms the structure of GO, which we will discuss later.

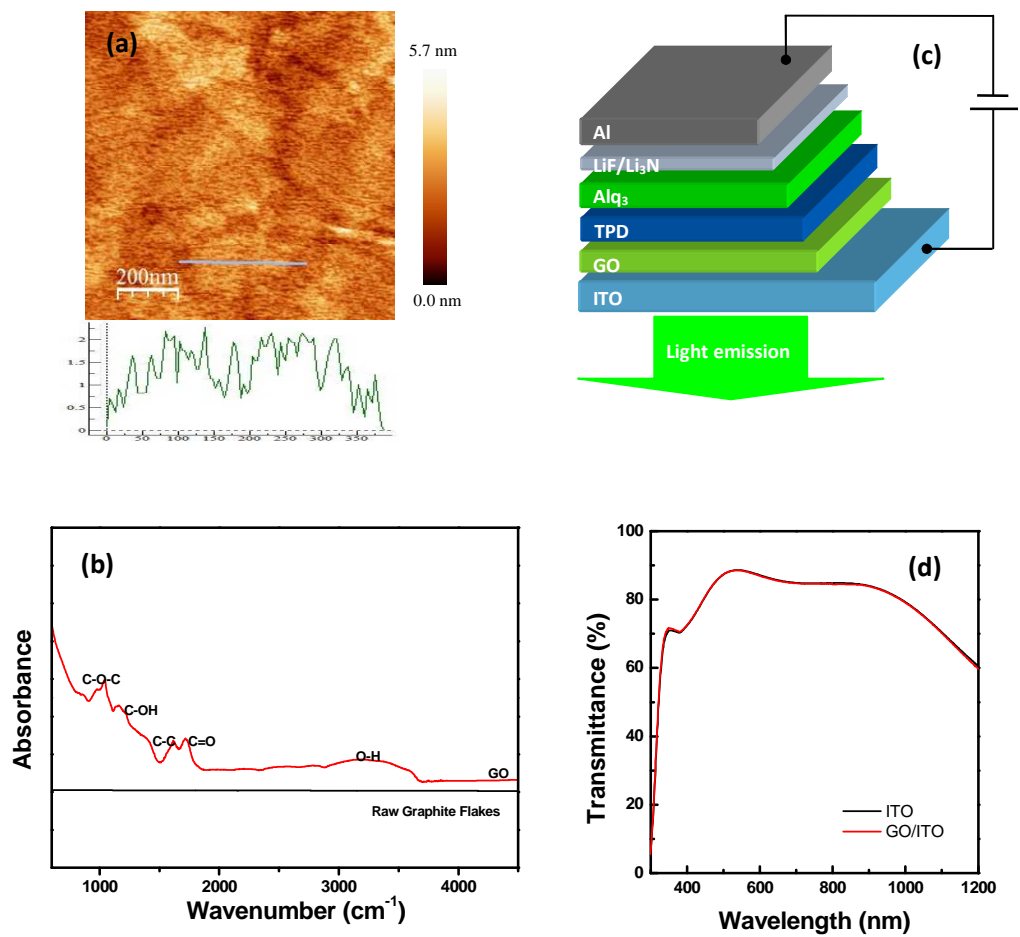

**Figure 1**. (a) AFM image of GO film spin-coated from 0.1 mg ml$^{-1}$ solution on silicon substrate (1×1 μm), (b) FTIR spectra for GO and raw graphite flakes, (c) Scheme of OLED device structure, (d) UV-vis for pure and modified ITO by 0.1 mg ml$^{-1}$ GO.

UV-vis shows that the transmittance is not significantly reduced from the reference with the GO interlayer on ITO substrate at low concentrations (Fig.1 (d)). The transmittances for both substrates are near-identical, especially in the visible region, and thereby not compromising the light emission of the OLEDs. When the concentration of GO is over 0.8 mg ml$^{-1}$ (not shown), the transmittance decreases significantly, which is not helpful in light management because of the lower optical out-coupling efficiency. Three lower concentrations

(0.02, 0.1, and 0.4 mg ml$^{-1}$) of GO are investigated as hole injection layers in OLEDs, and in Fig. 1 (c), the scheme for the device structure is shown.

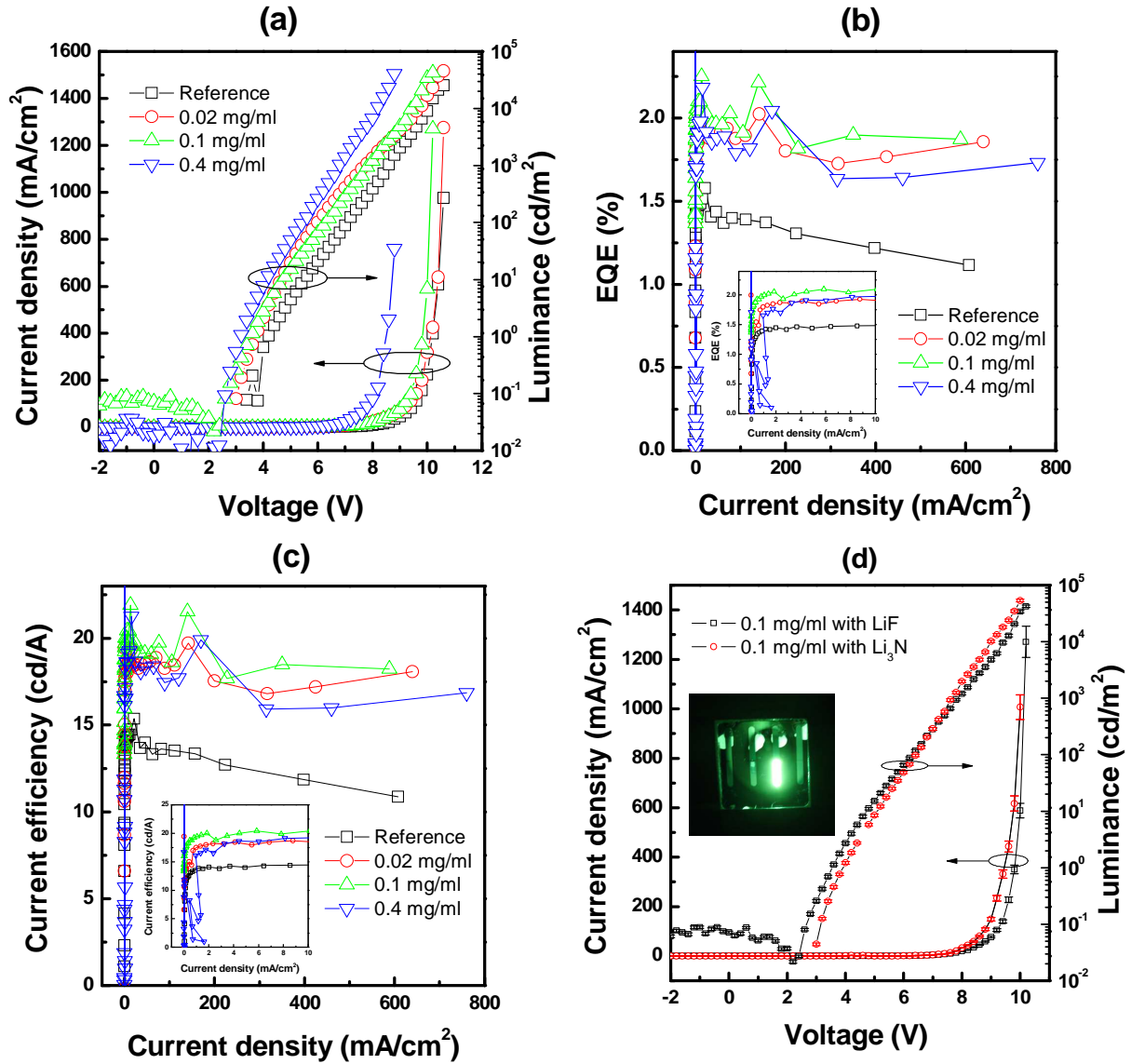

**Figure 2**. Device performance for OLEDs with pure and modified ITO by GO in different concentrations. (a) Current density-Voltage-Luminance, (b) EQE-Current density, (c) Current efficiency-Current density, and (d) The comparison of Current density-Voltage-Luminance characteristics for OLEDs on GO/ITO anode with different electron injection layers, inserted is a real-time picture during device operation.

The characteristics of the devices with GO interlayers are shown in Fig. 2 and the results for all the devices summarized in Table 1. The devices with GO interlayers work noticeably better than the reference device without GO in terms of turn-on voltage, luminance, external quantum efficiency (EQE) and current efficiency. For example, when 0.1 mg ml$^{-1}$ GO is used, the turn-on voltage is reduced from 4.2 V to 3.4 V, and the maximum luminance has a significant enhancement of more than 1.5 times. While the EQE has little improvement from 1.58% to 2.25 % because of the low outcoupling efficiency,[25] the current efficiency is increased from 15.36 cd/A to 21.90 cd/A, and the power efficiency reaches 4.07 lm/W from 2.26 lm/W, with nearly a two times improvement. The improvement in the device performance can be attributed to the functionality of the GO interlayer. Since all devices have the same structure on the cathode side, we deduce that the hole injection is improved by the GO interlayer based on the current-voltage characteristics (Fig. 2(a)). As the GO concentration increases, the turn-on voltage decreases, and current density increases, while luminance has little difference base on its already high value which has limited output based on the emitter material used in the OLED structure. This means that not all the current enhanced in the device contributes to light emission, with the device EL efficiency having an optimal value at a certain concentration. EQE and current efficiency reach optimal values at a moderate GO concentration of 0.1 mg ml$^{-1}$ for our OLED design architecture. The improvement of the luminance is significant compared with the reference device, for example, the maximum luminance is increased from 25,534 cd m$^{-2}$ to 45,368 cd m$^{-2}$. At 8.8 V when 0.4 mg ml$^{-1}$ GO is used, there is a sharp increase in the OLED output from 1,518 cd m$^{-2}$ to 40,785 cd m$^{-2}$, which is nearly 30 times enhancement. This is close to the limit of our OLED emitter structure, with further increase in luminance with current density increases at higher concentration of GO, such as 0.4 mg ml$^{-1}$, not being possible. A further possible reason is that there are not enough electrons to recombine with excess holes, which can be resolved by improving electron injection and transport strategy in the OLED design. Another reason results from the device degradation because of the very high current density generated in the device. It is known that aggregated joule thermal effects at high local electric field will damage devices.[26] We believe that the luminance can be improved further with appropriate design architecture including encapsulation and alleviation of the thermal effects during device operation.

Table 1. Summary of device performance with different ITO anodes.

| Anode [a] | Work function (eV) | Turn-on voltage (V) | Luminance (cd m$^{-2}$) | | EQE (%) | Current Efficiency (cd A$^{-1}$) | Power Efficiency (lm w$^{-1}$) |
|---|---|---|---|---|---|---|---|
| | | | Maximum | At 8.8 V | | | |
| ITO (LiF) | 4.6 | 4.2 | 25,534 | 1,518 | 1.58 | 15.36 | 2.26 |
| 0.02 mg ml$^{-1}$ GO/ITO (LiF) | 4.7 | 3.6 | 45,368 | 3,330 | 2.02 | 19.71 | 3.17 |
| 0.1 mg ml$^{-1}$ GO/ITO (LiF) | 4.9 | 3.4 | 42,377 | 3,572 | 2.25 | 21.90 | 4.07 |
| 0.4 mg ml$^{-1}$ GO/ITO (LiF) | 5.1 | 3.2 | 40,785 | 40,785 | 2.18 | 21.26 | 3.32 |
| 0.1 mg ml$^{-1}$ GO/ITO (Li$_3$N) | 4.9 | 3.6 | 53,635 | 6,570 | 2.27 | 22.13 | 2.91 |

[a]Content in parentheses means the electron injection layer used in the device.

Lithium nitride (Li$_3$N) is reported to provide better electron injection than LiF in OLEDs.[27] To get a more balanced hole and electron current and further improve our device performance, we use Li$_3$N as an electron injection layer to replace LiF. In Fig. 2(d), the comparison of device performance is shown for OLEDs with LiF and Li$_3$N, with the GO interlayer produced from the solution of 0.1 mg ml$^{-1}$. Device current is increased with Li$_3$N in place of LiF, which is attributed to the effect of Li$_3$N on electron injection. A luminance of over 53,000 cd m$^{-2}$ is now obtained compared with 42,377 cd m$^{-2}$ for the reference LiF device. A real-time device in operation with suitably high luminance is shown in Fig. 2 (d). Although improving the electron current may help provide efficient charge recombination and ensure higher luminance, the side effects of higher current density will compromise the device performance in terms of a reduced lifetime.

There are three further reports in the literature on the highest luminance for undoped $Alq_3$ fluorescent devices.[28-30] The most recent record is 127,600 cd m$^{-2}$ with all carrier Ohmic-contacts by the use of complicated p-doping technology and fullerene ($C_{60}$) contact with LiF/Al cathode.[28] A further device with an output luminance value of ~70,000 cd m$^{-2}$ with 2,9-Dimethyl-4,7-diphenyl-1,10-phenanthroline (BCP) in direct contact with LiF/Al cathode, and postpackaging annealing to improve this to ~90,000 cd m$^{-2}$ at a voltage of 15.5 V has been reported. But, the high luminance is obtained at relative high voltage; with the best value being less than 30,000 cd m$^{-2}$ at 10 V.[29] The third report on high luminance is an output of 54,000 cd m$^{-2}$ with optimized thickness of Ba/Al bilayer cathode, but Ba is very sensitive to the environment and therefore not ideal for real device applications.[30] For all the above reported records, devices are with encapsulation. Recently, we have reported a luminance of nearly 50,000 cd m$^{-2}$ with multi-walled carbon nanotube (MWCNTs) interlayer on ITO.[31] Compared with the reported best results, high luminance of over 53,000 cd m$^{-2}$ is obtained using our GO electrodes at a lower voltage, with most importantly a simple device fabrication process and without any device encapsulation or post processing. To our knowledge there are no reports with such high luminance OLEDs using the material GO, with the result positioned to unlock a route of applying GO in flexible OLEDs and other electrode applications.

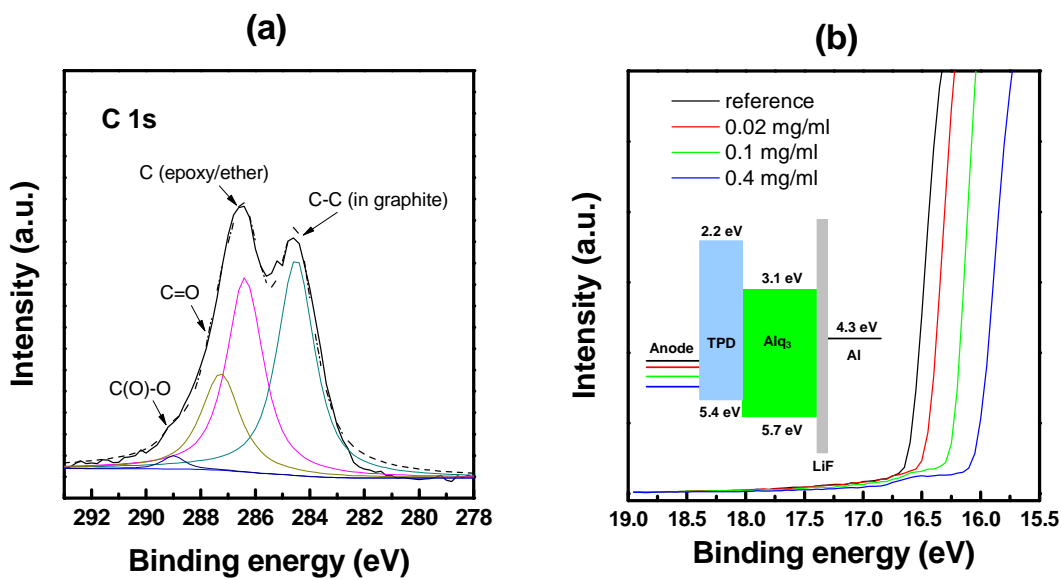

**Figure 3**. (a) Deconvoluted C 1s spectrum of GO film on ITO substrate with the concentration of 0.4 mg ml$^{-1}$. (b) The secondary electron cut-off region of the UPS spectra of different ITO anodes. Inset is the schematic energy level diagram of the device.

The influence of the GO interlayer on the effective WF of ITO was investigated by UPS and XPS. As shown in Fig. 3 (a), the C 1s spectrum of GO consists of three major components assigned to C-O (hydroxyl and epoxy, 286.4 eV), C=O (carbonyl, 287.9 eV), and C-(O)-O (carboxyl, 289.2 eV), and the peak at 284.8 eV attributed to C-C as in graphite.[32] The WF values are obtained by subtracting the secondary electron cutoff (SCO) with the He-I (21.2 eV) source used in UPS measurements. As evident in Fig. 3 (b), SCOs shift toward lower binding energy, indicating an increase in WF with GO interlayer on ITO. In particular, high concentration (0.4 mg ml$^{-1}$) GO can significantly increase the ITO's WF as compared to that of lower concentrations (Table 1). The high work function of GO are most likely due to the larger electronegativity of O atoms, which produce surface $C^{\delta+}$-$O^{\delta-}$ dipoles via extraction of electrons from graphene.[33] According to the proposed energy level diagram shown in the inset of Fig. 3 (b), the enhancement of current density for OLEDs with GO interlayers can be attributed to the improved effective WF and the creation of an interface energy step between ITO and TPD, leading to a reduced energy barrier and improvement of hole injection efficiency with concomitant luminance increase.

We have reported high luminance OLEDs with CNTs as the interlayer before,[31] but the effects of GO are very different from that of CNTs. CNTs are 1-dimensional tubular carbon materials that scatter charge due to the topological variations from a planar 2-dimentional surface and don't form an uniform film on substrates. Any improvement reported is due to the enhancement of local electric field effect of CNTs. While GO is 2-dimmensional material and can form an uniform film. As well known, GO is an insulator because of the disrupted $sp^2$ conjugation of the graphene lattice. However, the residual $sp^2$ clusters in GO can still allow for hole or electron transport to occur at the Fermi level by hopping espeicially when it is in contact with metal electrodes. In addition, It was reported that the large band gap of GO hinders transport of electrons from the cathode to the ITO, acting as an effective electron blocking layer, which will improve hole-electron recombination in the active layer.[17] Therefore the function of GO in OLEDs may be the change of work function, hole hopping near Fermi level and the higher hole-electron recombination efficiency.

In conclusion, we have demonstrated the application of solution-processable GO in OLEDs with a simple non-doped device structure. Our results demonstrate that higher luminance OLEDs with GO interlayer can be obtained by improvement of the electron injection strategy and appropriate device encapsulation. The results show a route to unlock a cheap and flexible ITO-free electrode system based on the material GO in flexible OLEDs and other plastic electronics.

**Acknowledgements**

This research was partly funded by a portfolio partnership award by the EPSRC, UK, and by E.ON AG, as part of the E.ON International Research Initiative. Responsibility for the content of this publication lies with the authors.